\documentclass[aps,prl]{revtex4}
\usepackage{subfigure,graphics} 
\usepackage{amsfonts,amssymb,amsmath}
\usepackage{flafter}
\usepackage{multirow}
\begin{document} 

\title{
Fermion stabilisation of extra dimensions and natural mass hierarchies} 
\author{Ian G. Moss}
\email{ian.moss@ncl.ac.uk}
\affiliation{School of Mathematics and Statistics, Newcastle University, NE1 7RU, UK}

\date{\today}


\begin{abstract}
Casimir forces from fermions in an extra dimension can stabilise 
Randall-Sundrum models. Contrary to previous expectations, the stabilisation
can produce large electroweak hierarchies without fine tuning. The Casimir stabilisation
occurs in the same five-dimensional fermion models which have been found to 
give the observed fermion mass hierarchies. This is shown for masses generated
by the Higgs mechanism and by a Higgsless symmetry breaking mechanism, 
where new light Kaluza-Klein states are presented

\end{abstract}
\pacs{PACS number(s): }

\maketitle

The standard model of particle physics comes with a number of mass hierarchies.
Fermion masses range from the top quark around $170\,{\rm GeV}$ down to the neutrinos at possibly
a few meV. If gravity is included in the standard model, then its magnitude is set by the Planck mass of 
around $2.5\times 10^{18}\,{\rm GeV}$, compared to the electroweak mass scale around $250\,{\rm GeV}$.
Finally, the small size of the cosmological constant is equivalent to a zero-level of energy in the vacuum 
with another scale of a few meV. 

The first two of these hierarchies have a natural explanation
in the Randall-Sundrum model, where the small size of the gravitational force is caused by
exponential decay of gravity across an extra dimension of spacetime spreading between two
surfaces, the ultra-violet scale (UV) and infra-red scale (IR) branes \cite{Randall:1999ee}.
These models also have the capacity to produce lepton and quark mass hierarchies
of comparable magnitude from fermions in the extra dimension (see e.g. \cite{Grossman:1999ra}).
 
One of the vital requirements of the Randall-Sundrum model is some kind of mechanism 
to fix the size of this extra dimension. The most successful is the Goldberger-Wise mechanism, 
which relies on the introduction of a five-dimensional scalar field \cite{Goldberger:1999uk}. 
In this paper we shall argue that the extra dimension is stabilised by the Casimir force between 
the two branes. The idea that the Casimir force may play a role is an old one, but it was discarded
because it seemed that the mechanism required an extreme degree of fine tuning
\cite{Candelas:1983ae,Garriga:2000jb,Garriga:2001ar,Flachi:2001pq,Flachi:2001bj,Flachi:2001ke}.
In fact, the mechanism turns out to be just as natural as the Goldberger-Wise 
mechanism. 

The Casimir stabilisation works for five-dimensional fermions with a particular
mass range. These turn out to be exactly the kind of five-dimensional fermions 
which have been found to give the correct quark mass hierarchies. This will be 
shown for quark masses given by both the conventional Higgs mechanism and by 
the topological symmetry breaking mechanism which is specific to higher dimensions. 
In the latter case, this will involve new light Kaluza-Klein states.

Before consideration of an alternative to the Goldberger-Wise mechanism, it is worth
considering why this mechanism works so well. The Randall-Sundrum model
is based on a five-dimensional geometry described by the metric
\begin{equation}
ds^2=e^{-2ky}\eta_{\mu\nu}dx^\mu dx^\nu+dy^2,
\end{equation}
with the UV brane at $y=0$ and the IR brane at $y=y_{IR}$. 
The Goldberger-Wise mechanism requires a massive bulk scalar field $\phi$. It will be convenient
to write the mass of the scalar in terms of the UV scale as $c_\phi k$. The field has potentials
on either brane, $V_{UV}(\phi)$ and $V_{IR}(\phi)$, whose detailed form need not 
concern us \cite{Goldberger:1999uk}. The scalar field settles into a ground state which depends on the
separation through the warp factor of the IR brane,
\begin{equation}
\sigma_{IR}=e^{-k y_{IR}}.
\end{equation}
When substituted back into the scalar field action, this gives the stabilising potential
$V(\sigma_{IR})$. The stabilisation mechanism has to somehow produce 
$\sigma_{IR}\sim 10^{-16}$, in order to obtain an IR scale
$k\sigma_{IR}\sim {\rm  1\ TeV}$. It turns out this can happen when the
bulk scalar mass satisfies
\begin{equation}
\epsilon=(4+c_\phi^2)^{1/2}-2\ll 1
\end{equation}
In this case, the potential has a `nearly conformal' form \cite{Konstandin:2011dr}
\begin{equation}
V(\sigma_{IR})\sim \delta V_{UV}+\sigma_{IR}^4P(\sigma_{IR}^\epsilon),
\label{gwpot}
\end{equation}
with a polynomial function $P$ depending on the various parameters
and a constant term $\delta V_{UV}$ which represents a shift in the tension of the
UV brane. The potential has a minimum at 
\begin{equation}
\sigma_{IR}\approx f_0^{1/\epsilon},
\end{equation}
where $P(f_0)=0$. Therefore very small values of $\sigma_{IR}$ are easily realised 
with very weak fine tuning of $\epsilon$. In this sense, the Goldberger-Wise 
mechanism produces a `natural' hierarchy. The only fine-tuned parameter is $\delta V_{UV}$, 
which is a manifestation of the unresolved cosmological constant problem.

The  alternative to the Goldberger-Wise mechanism is based on a generalisation of the
Casimir effect to five dimensions
\cite{Garriga:2000jb,Flachi:2001pq,Flachi:2001bj,Flachi:2001ke}. 
The original Casimir effect \cite{Casimir} is an attractive force between 
two conducting plates caused by quantum vacuum fluctuations. The
direction of the force is very much dependent on the geometrical set-up 
and the type of field, and Fermions in five dimensions with the simplest of boundary 
conditions give a repulsive force between the branes at small separations which can 
stabilise the extra dimension.

To see this in more detail, we can take the formula for the vacuum energy of a massive 
fermion given in Ref. \cite{Flachi:2001bj}. For a fermion of mass $ck$ in five dimensions, this is
\begin{equation}
V(\sigma_{IR})=\delta V_{UV}+\delta V_{IR}\sigma_{IR}^4+\Delta V(\sigma_{IR}),
\end{equation}
where $\delta V_{UV}$ and $\delta V_{IR}$ are quantum corrections to the tensions
of the respective branes,
\begin{equation}
\Delta V(\sigma_{IR})=-{(k\sigma_{IR})^4\over 16\pi^2}\int_0^{y_{IR}}dy\,
y^3\ln\left\{1-{K_{\epsilon/2}(y)I_{\epsilon/2}(\sigma_{IR}y)\over 
K_{\epsilon/2}(\sigma_{IR}y)I_{\epsilon/2}(y)}\right\}
\end{equation}
and $\epsilon=2c-1$. Results for the massless fermion, $c=0$, can be reduced 
to elementary functions and this case has accordingly received most attention,
but it does not lead to naturally small values of $\sigma_{IR}$ \cite{Flachi:2001bj}. We shall focus
instead on the small $\epsilon$ case which has seemingly been overlooked. 
Numerical evaluation of the integral for small values of $\sigma_{IR}$
shows that the potential has the Goldberger-wise form (\ref{gwpot}),
\begin{equation}
V(\sigma_{IR})\sim \delta V_{UV}+\delta V_{IR}\sigma_{IR}^4+C\sigma_{IR}^{4+|\epsilon|},
\label{veff}
\end{equation} 
where $C\approx 1.3\times10^{-3}k^4$. The branes undergo a similar natural stabilisation with 
minor fine tuning of $\epsilon$ when $\delta V_{IR}<0$. We will assume from now on $\delta V_{IR}$ 
takes the correct sign. 

If there are multiple fermion fields, then the one with the smallest value 
of $\epsilon$ has the largest effect on the potential for small $\sigma_{IR}$.
Gauge-bosons can also contribute to the Casimir force, but these have
integer $\epsilon$, like the massless fermions. The hierarchy is set by the mass of one of 
the fermion fields.

\subsection{Naturalness and fermion masses}

We turn next to the fermion mass hierarchies (following e.g. \cite{Grossman:1999ra}). 
We shall see that the same parameter
ranges that lead to these hierarchies are the same ones which are needed for natural
stabilisation of the extra dimension.
For simplicity we shall take two quark fields ${\cal Q}$  and ${\cal U}$ in five dimensions. 
The pair will reduce to the left and right handed components of a single quark field in four 
dimensions, which may be the top quark, for example. The masses of the fields
are taken to be $\pm ck$, for some constant $c$. This signs of the mass terms determine
near which brane the fermion field takes its largest values.

The fermion fields decompose into Kaluza-Klein modes  $u^{(n)}_L(y)$ and $u^{(n)}_R(y)$,
for the left and right components of ${\cal Q}$ (i.e. $\gamma_5{\cal Q}_L=-{\cal Q}_L$),
\begin{align}
{\cal Q}&=\sqrt{k}
\sum_n e^{2ky}\left\{u_L^{(n)}{\cal Q}_L^{(n)}+u_R^{(n)}{\cal Q}_R^{(n)}\right\}\\
{\cal U}&=\sqrt{k}
\sum_n e^{2ky}\left\{u_R^{(n)}{\cal U}_L^{(n)}+u_L^{(n)}{\cal U}_R^{(n)}\right\}
\end{align}
Modes switch around due to the five-dimensional mass convention. The four-dimensional
theory is canonically normalised if we normalise the modes using a weight factor $e^{ky}$.

The eigenvalues of the modes determine the Kaluza-Klein masses $m^{(n)}$. 
These Kaluza-Klein modes are too heavy to correspond to standard model fermions, which must
therefore be represented by the massless modes. For boundary conditions 
${\cal Q}_R={\cal U}_L=0$ on the branes, the zero modes are
\begin{equation}
u_L^{(0)}(y)=\left({2c-1\over 1-\sigma_{IR}^{2c-1}}\right)^{1/2}e^{-cky},\label{massless}
\end{equation}
when the modes are properly normalised.

The conventional way for the massless modes to acquire mass is through a Higgs field ${\cal H}$ 
confined on the IR brane, with Yukawa terms also defined on the IR brane,
\begin{equation}
{\cal L}_Y=\sqrt{g}k^{-1}y_Q\,{\cal H}\overline{\cal Q}{\cal U}.
\end{equation}
 The Yukawa coupling $y_Q$ defined this way is dimensionless. The 
 Yukawa coupling strength $y_q$ of the canonically normalised fields ${\cal Q}_L^{(0)}$ and
 ${\cal U}_R^{(0)}$ is obtained by substituting the zero modes (\ref{massless}). We also have 
 to take into account the fact that the canonically
 normalised Higgs field is $H={\cal H}/\sigma_{IR}$. This gives
 \begin{equation}
 y_q=y_Q{\epsilon \,\sigma_{IR}^{\epsilon}\over  1-\sigma_{IR}^{\epsilon}}
 \end{equation}
 where $\epsilon=2c-1$, as before. Since $\sigma_{IR}\sim 10^{-16}$, a wide range
 of Yukawa couplings can be obtained from a modest range of values for $\epsilon$. 
 This has been exploited in phenomenological models 
 \cite{Casagrande:2008hr,Morrissey:2009tf}, which include
 multiple generations of quarks and leptons. These  can reproduce the wide fermion mass range found
 in the standard model. 
  
 The significant feature of the fermion masses from our new perspective is that the largest 
 quark mass requires a relatively small value of  $\epsilon$, which is exactly is what we need for 
 natural stabilisation of the brane separation from the fermion vacuum energy. In the model of 
 Ref \cite{Casagrande:2008hr} for example, $c_{\cal Q}=0.521$ for the top quark doublet.
 The minimum of the potential ({\ref{veff}) is then at 
 $\sigma_{IR}\approx (-C/\delta V_{IR})^{25}$ and will reproduce the electroweak hierarchy without 
 any need for fine tuning.

\subsection{Higgsless models}

In five dimensions there is an alternative form of Higgs mechanism,
where the symmetry is broken by a non-zero gauge field stretching between 
the two  branes \cite{Hosotani:1983,Hosotani:1989,Flachi:2001bj,Burdman:2002se}. 
The gauge field can be removed locally by gauge transformations, but the holonomy 
remains in the form of non-trivial boundary conditions on the fermion fields. In the Randal-Sundrum 
context, this type of breaking goes under many names, and we shall follow \cite{Flachi:2001bj}
and refer to this as topological symmetry breaking. An interesting feature of these models
is that the symmetry breaking parameters  develop potentials {\it and} kinetic terms from quantum corrections 
to the effective action \cite{Moss:2002dv}. 

We shall see now how fermion masses obtained from topological symmetry breaking
are consistent with the natural stabilisation of the extra dimension, in the context of
a simple toy model which brings out the essential features. Following Ref \cite{Flachi:2001bj},
we can take a doublet of fermions in five dimensions and impose boundary conditions
on the UV and IR branes which respect reflection symmetry `up to holonomy',
\begin{equation}
\begin{pmatrix}
{\cal Q}\\ {\cal U}\\
\end{pmatrix}=
-\tau^3\gamma_5
\begin{pmatrix}
{\cal Q}\\ {\cal U}\\
\end{pmatrix}
\hbox{ UV},
\quad
\begin{pmatrix}
{\cal Q}\\ {\cal U}\\
\end{pmatrix}=
-\tau_\theta\gamma_5
\begin{pmatrix}
{\cal Q}\\ {\cal U}\\
\end{pmatrix}\hbox{ IR},
\end{equation}
where $\tau_\theta=\tau^3\cos 2\theta+\tau^1\sin 2\theta $ and $\tau^i$ are the Pauli matrices.  
We shall also take the fermion mass matrix to be $ck \tau^3$. (With this convention, the fermions fit naturally
into the eight component spinor representation of the full Lorenz group in five dimensions.
We would require extra mass terms or additional fermions to break chiral symmetry.)
 
Zero modes only exist for $\theta=0$, and have the same form as before (\ref{massless}). 
These combine to give a singe massless fermion in four dimensions. When the symmetry 
breaking is turned on, the massless modes disappear, but we shall discover that one of the 
Kaluza-Klein modes becomes exceptionally light, giving a natural hierarchy between the 
quark or lepton mass scale and the Kaluza-Klein mass scale.

To see how this comes about, we need the explicit forms of the fermion mode functions
given, for example, in \cite{Flachi:2001bj},
\begin{align}
u_L&=a_L f_L+b_Lg_L\\
u_R&=a_R f_R+b_Rg_R
\end{align}
where $f_{L,R}=e^{ky/2}J_{c\pm1/2}(m^{(n)}e^{ky}/k)$ and 
$g_{L,R}=e^{ky/2}J_{-c\mp1/2}(m^{(n)}e^{ky}/k)$ up to normalisation factors. 
(Note that there are different coefficients for the
the ${\cal Q}$ and ${\cal U}$ components.)
The Kaluza-Klein masses are given by inserting the mode expansions into the boundary
conditions and looking for non-trivial solutions. This leads to
\begin{equation}
{g_R(y_{IR})f_R(0)-f_R(y_{IR})g_R(0)\over g_L(y_{IR})f_R(0)+f_L(y_{IR})g_R(0)}
=\pm\tan\theta.
\end{equation} 
In the regime of interest, $\sigma_{IR}\ll 1$, there is a tower of Kaluza-Klein fermions
with masses given by $m^{(n)}\approx j_n\,k\sigma_{IR}$, where $j_n$ are the zeros
of the Bessel function combinations $\cos\theta f_R(y_{IR})\mp \sin\theta f_L(y_{IR})$. 
This mass splitting for the $\pm$ options reflects the breaking of the $SU(2)$ symmetry of the 
Kaluza-Klein tower by the boundary conditions.

Of particular significance is the existence of a single light fermion which is separated
from the rest of the Kaluza-Klein tower. The mass of this fermion can be found by taking
small argument expansions of the Bessel functions,
\begin{equation}
m^{(0)}\approx (c-1/2)k\tan\theta\,\sigma_{IR}^{2c}.
\end{equation} 
The existence of this light fermion is, of course, related to the fact that there is a massless
fermion when $\theta=0$. Note, however, that the mass is small even when $\theta\sim 1$
due to the small size of $\sigma_{IR}$. The new state is similar to the one found in
Ref. \cite{Csaki:2003sh}, but in that work the state depended on introducing a 
mass-like term on the boundary branes.

We can compare the light fermion mass to the IR scale $k\sigma_{IR}$,
\begin{equation}
{m^{(0)}\over k\sigma_{IR}}\approx \epsilon \tan\theta\,\sigma_{IR}^{\epsilon}.
\end{equation}
where $\epsilon=2c-1$ as before. A value of $\epsilon\sim0.1$, corresponding to the standard 
model fermion with largest mass, would 
provide the natural stabilisation of the brane separation by the fermion vacuum energy.
With the small value of $\sigma_{IR}$, different values of $\epsilon$
would then give natural fermion mass hierarchies.

In conclusion, we have seen that the fermions which lead to natural mass hierarchies
in the context of the Randal-Sundrum model also lead to natural stabilisation of the
extra dimension through the Casimir force. The argument is quite general, and it would be interesting to 
see how it plays out in phenomenological models. The stabilisation mechanism still contains
a residual ambiguity because the renormalised brane tensions are not fixed
by model parameters accessible to experiment. It may, however, be possible to relate the renormalised 
brane tensions to the supersymmetry breaking scale in supersymmetric models,
along the lines of Ref. \cite{Fabinger:2000jd}. In comparison, the alternative 
Goldberger-Wise stabilisation mechanism has at least five parameters that determine
the brane separation but are otherwise hidden from observation.

Like the Goldberger-Wise mechanism, the model does not solve the cosmological constant problem, 
so that one brane tension has to be fine tuned to make the cosmological constant vanish.
There are also difficulties in the construction of plausible cosmological scenarios in the Randal-Sundrum
context. Nevertheless, the simplicity of the Randal-Sundrum scenario when stabilised
by standard model fields alone is worth persuing.
 
\section*{Acknowledgements} 
The author would like to thank the STFC for financial support 
(Consolidated Grant ST/J000426/1).

\bibliography{paper.bib}

\end{document}